# Frustrated Total Internal Reflection: Resonant and Negative Goos-Hänchen Shifts in Microwave Regime

## Min Qu[*], Zhi-Xun Huang

*Department of Communication Engineering, Communication University of China, Beijing 100024,China*

**Abstract**

It is well-known that the variations of Goos-Hänchen shifts (GHSs) are closely associated with the enengy-flux provided by evanescent states in the case of total internal reflection. However, when the frustrated internal total reflection (FTIR) is realized with a Polymethyl Methacrylate (PMMA) double-prism system operated in the microwave frequency range of 8.2 $GHz$ to 12.4 $GHz$, we observe that the GH shifts for the reflected beam show periodic resonances with varying the operation frequency or the air layer thickness, which is different from the variation of the corresponding reflected energy. Moreover, in another FTIR based system introduced by a composite absorptive material slab with a two-dimensional top layer of frequency selective surface (FSS), the GHSs for reflected beam are discovered as not only resonant but also negative with the incidence of transverse electric that is TE polarized.

*Keywords:* FTIR, resonant GH shifts, negative GH shifts, energy-flux

## 1. Introduction

The phenomenon of frustrated total internal reflection (FTIR) with a dynamic range of altering reflection, which was firstly found by Newton, has played a critical role in optical experiment and engineering, such as multi-touch sensor, single-pulse Q switching in cavity resonator and single - fiber breakdown spectroscopy [1,2]. FTIR surprisingly makes the Goos-Hänchen shifts (GHSs) of the reflected beam present positive, negative or even zero since the reflected energy-flux is changed. In contrast with the total internal reflection (TIR), the significant difference is that, more or less, the energy from evanescent states in FTIR would flow across a separating medium (medium 2) and enter into a region occupied by a higher index of refraction material (medium 3) rather than totally coming back into medium 1.

The multilayer configuration of FTIR is a fine bridge between micro and macro world. Tremendous studies have been carried out on this topic, which covers a variety of areas including nano-optics, microwave optics, optical fibre communication, biosensor technology and so on. In theory, Ghatak et

---
[*]Corresponding authors. Tel.: +86 134 262 328 78 (Min Qu).

E-mail addresses: lydiaqu09@gmail.com ( Min Qu),

huangzhixun@gmail.com (Z.X. Huang).

al. [3], Hsue and Tamir [4] give analytical expressions for reflected and transmitted GHSs based on the standard approach of stationary phase. Their results show that the GHSs vary rapidly with the frustrating layer thickness. However, Zhu et al. [5] proposed a different expression for transmitted GHSs under partial reflection, which presents a resonant relation between GHSs and the air layer thickness, also under the stationary-phase framework. In recent years, as the new elements (absorbing media or active media) constantly being introduced, the limitation of stationary-phase method becomes more and more distinct. Consequently, some people choose the straightforward calculations of light moments theory [6] or Renard's [7] energy-flux theory. As to the experimental work, at present there are two focuses: (1) how to enhance the GHSs considerably [8, 9] and (2) how to implement the negative shifts [10]. In fact, for negative GHSs, instead of reporting any experimental data, most studies [11, 12] solely focus on the possibility for the negative GHS from the perspective of the pure theory.

In this paper, we propose two systems under FTIR. The first system includes a Polymethyl Methacrylate (PMMA) double-prism operated in the microwave frequency from 8.2 $GHz$ to 12.4 $GHz$. The second microwave system contains a configuration of PMMA prism-air-composite metal sheet with frequency selective surface (FSS). The investigations of the two systems reveal the behaviors of both the resonant and the negative GHSs for specific beam and polarization, which are described in detail in the next several sections.

## 2. Experimental Arrangement

The first system is illustrated in Fig. 1(a). In the transmitter part, the microwaves, generated by Agilent MXG-N5183A Signal Generator with the amplitude set to 13 $dBm$, are fed into a parabolic antenna with diameter $D = 200mm$ to create a circular Gaussian beam. This beam travels through an absorptive material with a central rectangle aperture of $100 \times 100\,mm$, as shown in Fig. 1(b), and immediately penetrates into a symmetrical system, resembling most theoretical models that consist of two same prisms. Both prisms are made by PMMA with a dimension of $320 \times 320 \times 308\,mm$ and have a refraction index with measuring value of approximate $n = 1.605$ (corresponding $\theta_c = 38.5^o$) at the frequency of 10 $GHz$ in the system. In the receiver part, the reflection signals is detected by a standard gain horn with an aperture of $60 \times 84$ $mm$ which is connected to Agilent E4407b Spectrum Analyzer.

Fig. 1. The pictures of the first system: (a) the overall view of the system for measuring evanescent states and positive GHSs; (b) the setup illustrating absorbing materials;

For the second system intended for the observation of the negative GH shift, the multilayer structure under test is shown in Fig. 2(a). There, the composite material slab with a two-dimensional top layer of frequency selective surface (FSS) is separated by an air at a distance of about $\lambda/10$ from the PMMA prism. These setups, appear to be similar to the Otto device used to excite surface plasmon polaritons. However significant modifications have been made in our investigations, as described in Fig. 2.

In addition, a mirror polished aluminum sheet clung to the inclined plane of the first prism is used to find the position of geometrical reflected beam.

Fig. 2. The structure of the second system: (a) the air lies between the prism and the FSS surface; (b) the photo of the setup used to realize negative GHSs.

# 3. Experiments and Results

## 3.1. Transmission and Reflection

In the preliminary experimental procedures, after the completion system alignment, instrument calibration, absorber arrangement, beam polarization adjustment and refraction index determination, it is necessary to measure the transmission and reflection damping as varying the air layer thickness $d$ with the variation range of $1\sim36$ mm in order to determine the dynamic range of FTIR. The schematic diagram of microwave damping measurement is given in Fig.3. If $0\,dB$ was set at the field points near the parabolic aperture, the damping values we have measured would not be more than 40 $dB$.

Fig.4 shows the data fitting results from measurement does not always decrease linearly, overall trends are, more or less, in agreement with Salomon's experimental results at optical frequency accomplished in 1991 [2]. The case where the width of the air gap between the two prisms is smaller than a threshold value such as 9.5 mm at 10.5 $GHz$, which is approximately corresponding transmission depth of evanescent states, the amplitude of the transmission field through the second prism decreasing quadratically. The TIR condition is also no longer valid. Whereas the case where the two prisms are sufficiently far from each other and the energy that tunnels through the potential barrier of the air gap decreases exponentially, is referred to as the quantum tunneling regime. Furthermore, the reflection power in the FTIR regime of $d<10mm$ ( $f=10.5GHz$ ) increases rapidly with the air gap thickness, while in quantum tunneling regime of $d>10mm$ the energy slowly approach to saturation, see in Fig.5. These results, in accord with the theoretical results in Ref.[3][4], prove that the energy-flux in the FTIR structure of double-prism complys with the conservation of energy.

Fig.3. The schematic diagram of microwave damping measurement.

Fig.4. The intensity transmitted in the air gap as a function of the air gap thickness. The result is based on the conditions of TM polarization, beam width of 100×100 mm and $n=1.605$ .

Fig.5. The intensity reflected in the air gap as a function of the air gap thickness. The result is based on the conditions of $f=10.5GHz$ ,TM polarization, beam width of 100×100 mm and $n=1.605$ ).

## 3.2. The investigation GHSs

According to Renard's model, the reflected GHSs at a single interface or at a multilayered structure depend on the Poynting vector in the totally reflecting medium. However, the upper curve in Fig.6 shows when the second prism (medium 3) lies to the distance of $d < 12mm$ ( $f = 10GHz$ ) from the first prism (medium 1), the GHSs for the reflected beam show periodic resonances with varying the air layer thickness. Comparing the upper curve with the below curve in Fig.6, we can see that the variation of GHSs does not follow the trend of its corresponding reflected energy in the FTIR regime. Additionally, as varying the operation frequency and keeping the parameter $d = 4mm$ , we also see the periodically resonant GHSs, which is generally smaller than the corresponding ones in the case of TIR, as shown in Fig. 7.

Fig.6. The comparison diagram between reflected GH shifts and reflected energy versus the air gap thickness The result is based on the conditions of $f = 10GHz$, TM polarization, beam width of 100×100 mm, the incident angle of 45º and $n = 1.605$.

Fig.7. The reflected GHSs versus frequency with the air gap of 4mm. The result is based on the conditions of $f = 10GHz$ ,TM polarization, beam width of 100×100 mm, the incident angle of 45º and $n = 1.605$.

To further prove the constitutive relation between the GHSs and the reflected enengy-flux, we set an air gap of 20 mm and examine GHSs. Fig. 8 shows the GHSs agree well with those of TIR at a single interface. It implies that, when the two prisms weakly coupled with each other, the TIR is not, or at least not much, perturbed.

Fig.8. The reflected GHSs versus frequency for two different cases. The result is based on the conditions of $f = 10GHz$ ,TM polarization, beam width of 100×100 mm, the incident angle of 45º and $n = 1.605$.

## 3.3. Novel implement of Negative GH shifts

Like the photonic crystal with photonic bandgap, the FSS structure consists of electromagnetic bandgap (EBG). The transmission capacity of the metal FSS is weaker than would be expected from the pure metal. When falling into the higher frequency band of our system, the FSS slab presents prominent absorption, as shown in Fig. 9. The ratio between the lattice period of the metal FSS and the wavelength (inside the air) is slightly less than 1/2 (about 0.49). Thus the zeroth reflection would not appear.

Fig.9. The comparison diagram of reflected energy between Al slab and FSS slab. The result is based on the conditions of TE polarization, beam width of 100mm×100mm, the incident angle of 45º and $n = 1.605$.

As well known, in optical frequency, large negative GHSs can be generated on a metal surface [11]. However, we realized not only negative but also resonant GHSs for TE polarized wave on the EBG metal surface, as indicated in Fig. 10. On the other hand, for TM polarized wave we only observe the positive GHSs when the other conditions remain the same.

Fig.10. The negative GH shifts versus frequency. The result is based on the conditions of TE polarization, beam width of 100mm×100mm, the air gap of 2 mm, the incident angle of $45^o$ and $n = 1.605$).

## 4. Conclusion

In conclusion, we have measured the transmission / reflection energy-flux and GHSs upon two FTIR systems in the microwave regime. We found that the resonant GH shifts for reflection beam always exist in such FTIR structures, whether with a homogeneous dielectric prism or a composite FSS slab. In fact, it is reported that the Febry-Pérot resonances at a interface may be excited and can dramatically enhance the resonant GHSs of light and has been widely used in the integrated optics and optical sensors [14]. Thus, the structures in this article may have potential applications in the design of microwave device and microwave optics.

What is more, under our expectation, also negative GH shifts were observed at a two-dimensional FSS slab interface. For the future it is expected, that a pertinent quantum theoretical evidence could be used to explain the formation of negative shift.

## Acknowledgement

The author would like to appreciate Prof. G. Z. Lu for his help in establishing the experimental systems. The study was financially supported by the Chinese National Science Fund under Grant No. 200800330002 and the Scientific Research Bureau of Communication university of China under Grant No.XNG0906. We also gratefully acknowledge the generous help from Guang Zhou Antas Chemical Corporation.

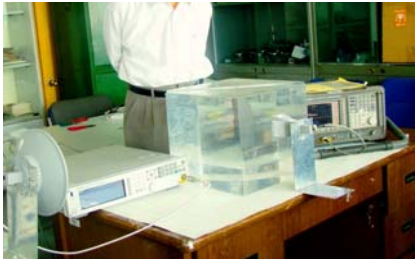

(a)

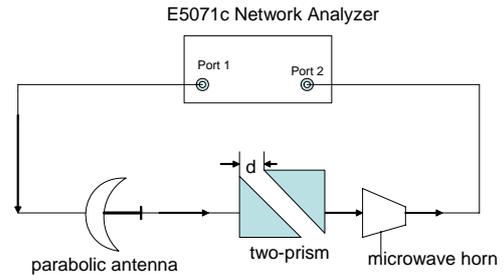

Fig.3.

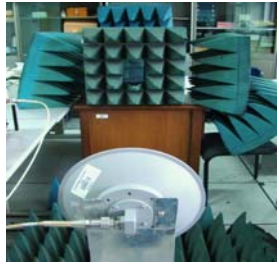

(b)

Fig. 1.

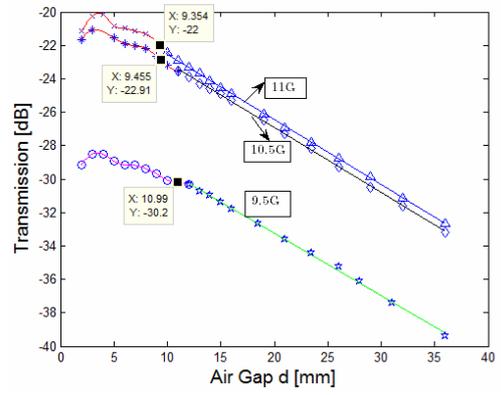

Fig.4.

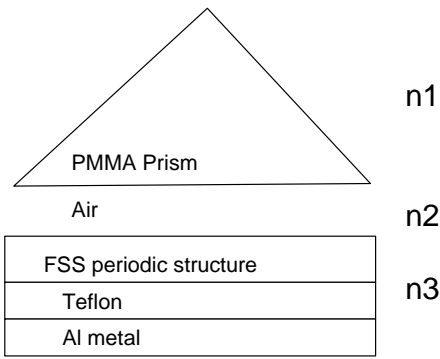

(a)

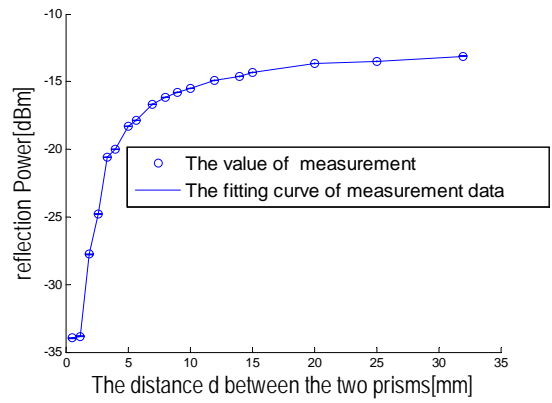

Fig.5.

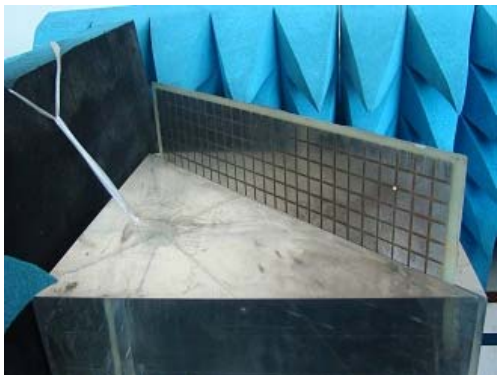

(b)

Fig. 2.

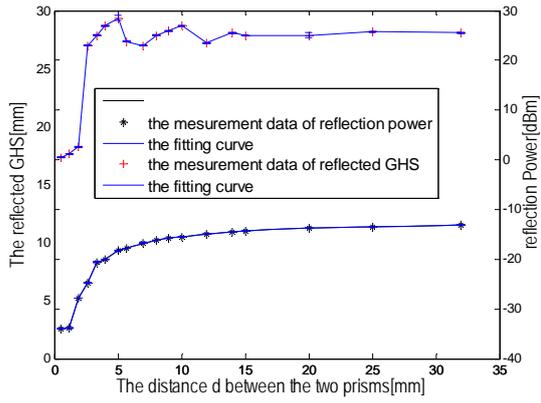

Fig.6.

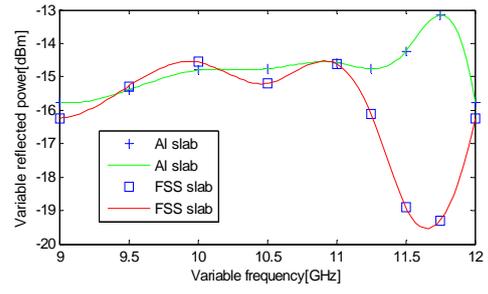

Fig.9.

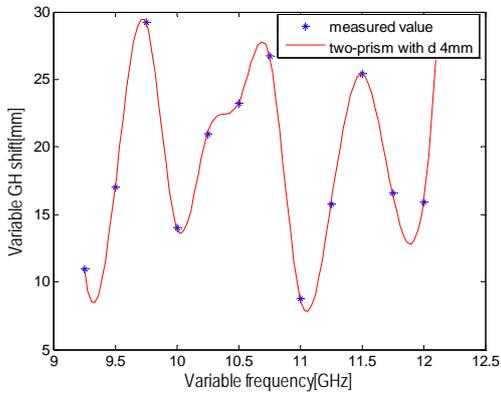

Fig.7.

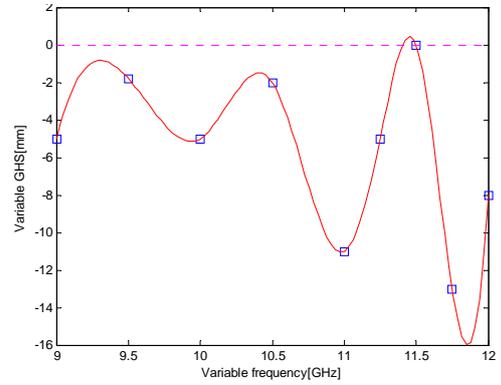

Fig.10.

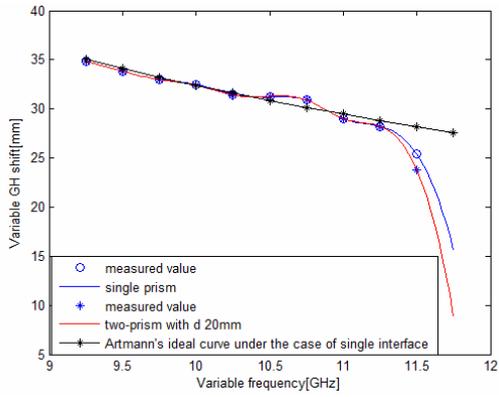

Fig.8.